\documentclass[preprint2]{aastex}
\slugcomment{submitted to {\it The Astronomical Journal}}
\begin{document}
\title{Field Stars, Open Clusters, and the Galactic Abundance Gradient}
\author{Stuartt Corder and Bruce A. Twarog}
\affil{Department of Physics and Astronomy, University of Kansas, 
Lawrence, KS 66045-2151}
\affil{Electronic mail: corder@kustar.phsx.ukans.edu, 
twarog@ukans.edu}
\begin{abstract}
The effects of the orbital diffusion of stars and clusters on the galactic
abundance gradient have been modelled. For a single generation of field stars, 
the results confirm previous analyses based upon the assumption of a linear 
gradient with galactocentric distance. Though the dispersion in metallicity 
at a given radius grows with time, the underlying slope across the disk 
remains unchanged. For a disk dominated by a discontinuity in [Fe/H] 
beyond the solar circle, but no gradient within the inner or outer disk, a
gradient is derived for any sample whose galactocentric baseline
overlaps the discontinuity. The narrower the range in distance around the
discontinuity, the steeper the gradient. The consequence of orbital diffusion 
is to shift the observable edge of the transition in [Fe/H] toward 
smaller galactocentric
distance, enhancing the detection of a gradient. In contrast, star clusters of
a single generation evaporate so quickly that the small surviving sample of the
oldest clusters is statistically incapable of defining any gradient except a
linear case.  Inclusion of multiple generations of stars and clusters with a 
plausible age-metallicity relation enhances these patterns by expanding the
metallicity range at a given galactocentric distance through the inclusion of
stars with a wide range in both age and galactocentric origin. For field stars,
the metallicity distribution of stars in the solar neighborhood is virtually
indistinguishable for the two cases modelled. Because ages, abundances,
and distances can
be reliably determined for clusters and any sample of clusters is invariably
dominated by those clusters formed over the last 2 Gyrs, the cluster population
offers a viable means for finding detailed structure within
the recent galactic abundance gradient. Using clusters to extend the
galactocentric baseline would supply an additional critical test of the
discontinuity. 
\end{abstract}
\keywords{Galaxy: evolution -- Galaxy: disk -- Galaxy: kinematics and dynamics}
\section{INTRODUCTION}
The fundamental goal of Galactic stellar populations is an understanding
of the chemical and dynamical history of the Milky Way.
Since the seminal work of \citet*{ELS62},
the implicit assumption that kinematics, composition, location, and age
were correlated has gradually given way to a more chaotic view where modest
trends among these parameters are subsumed beneath a large dispersion
in each as a function of time. Though a fragmented origin for the Milky
Way is most commonly associated with the halo, as first suggested by
\citet*{SZ78}, the need for a greater degree of
randomness within the populations has been pushed toward more
recent galactic epochs, in part, by the evidence for 
and analyses of the thick disk
population \citep*{GR83} and the field star analysis of the solar neighborhood
by \citet{E93}.

The work of \citet{E93} has taken on particular significance because of 
its implication that the
metallicity of stars at a given age covers a range of at least a factor of
four and possibly as large as a factor of eight, significantly larger
than found in earlier analyses of solar neighborhood stars \citep*{TW80,ME91}. 
Paradoxically, the increased scatter appeared 
in a sample which had supposedly more reliable ages 
and abundances, while reproducing the same overall mean trends with age
as the earlier studies. Though \citet{E93} explicitly
stated that their sample was biased and should not be used to analyze the
age-metallicity history of the disk, their own analysis and  discussion
of the data implied that the bias was modest and that the metallicity
spread was real. The bias originated in the sample selection. The distribution
of [Fe/H] among the bright F stars has long been known to exhibit a
paucity of stars with [Fe/H] below --0.4, supposedly the metallicity
range populated predominantly by the older stars of the disk. Since the
goals of \citet{E93} extended beyond the delineation of the age-metallicity 
relation to any metallicity-based trends, the apparent-magnitude-limited
sample was expanded to include a significant number of F stars with
lower [Fe/H] to ensure an approximately uniform distribution between 
[Fe/H] = +0.2 and --0.9. Fortunately, with proper weighting of the sample
to correct for the distortion, the bias can be and was, in theory, 
accounted for.  The other potential source of sample distortion, orbital
diffusion, was supposedly eliminated by the kinematic analysis, i.e., the stars
of a given age that formed within an annulus of modest width centered on 
the sun exhibited the same degree of abundance scatter as the overall sample. 
Attempts to explain
the degree of inhomogeneity within the context of galactic
evolution have been numerous \citep*{WFD96, PE96, VDH97}.

Beyond the lack of an explanation for the absence of 
comparable scatter in earlier
field star analyses of an unbiased sample, the conclusions of \citet{E93} are
contradicted by two observational constraints from a larger scale than
the solar neighborhood: 

(a) Studies of HII regions across the galactic
disk indicate that at a given galactocentric distance, the dispersion in
[m/H] is on the order of $\pm$ 0.1 dex, significantly smaller than found
within the field star sample, though there is evidence of a gradient in
[m/H] of typically between --0.05 and --0.10 dex/kpc \citep[see][and
references therein]{HW99}.

(b) \citet*{TAT97} have shown that the open 
clusters within galactocentric distance ($R_{GC}$) of 10 kpc ($R_{GC}$ for
the sun is 8.5 kpc), when placed on a uniform metallicity scale,
have a mean [Fe/H] $\sim$ 0.0 with a dispersion of only $\pm$ 0.10 dex.
This result occurs whether one uses all the clusters or separates them into
young and old (age greater than 1.5 Gyr). No cluster within $R_{GC}$ = 10 kpc 
has a reliably determined [Fe/H] below --0.2, while the majority of
the field star sample, adjusted for selection effects, populates this
regime.
                              
There are a variety of ways of explaining these discrepancies. 
The HII regions could be discarded because
they represent the current state of the interstellar medium. They would
be irrelevant if the chemical homogeneity of the gas within the disk
at a given galactocentric distance today is not typical of the dispersion
in [Fe/H] 2 to 10 Gyrs ago or if the instantaneous dispersion is not
representative of the spread after a few Gyrs worth of differential 
galactic rotation.
The former explanation implies that we somehow live at a special time in
the history of the galactic disk and is contradicted by the open cluster
data to an age of approximately 5 Gyr. The latter interpretation 
implies some form of orbital diffusion.
Stars formed at different galactocentric distances, within the
context of a radially-dependent [Fe/H], are dispersed with time over
increasingly larger $R_{GC}$, turning the range in the abundance with
$R_{GC}$ into a spread in [Fe/H] at a given $R_{GC}$ \citep{WFD96}. 
This interpretation
is challenged by the kinematic analysis of \citet{E93} discussed above and by
the open cluster data which exhibit little to no evidence for a gradient
in [Fe/H] between $R_{GC}$ = 6 and 10 kpc. There is, however, a significant
discontinuity in [Fe/H] for clusters beyond $R_{GC}$ = 10 kpc; the
mean abundance for the modest sample beyond this transition is
[Fe/H] $\sim$ --0.3.   

For the open cluster sample, the only objection possible is the rather
artificial one that the sample is unrepresentative of the field star
population. Either the open clusters do not form with the same abundance
distribution as the field stars or the metal-poor portion of the cluster
sample is preferentially destroyed. Any attempt to use abundance errors
to explain the result is contradicted by the care used in selecting, analyzing,
and combining the observational data, as well as the simple point that the
dispersion of $\pm$0.10 in [Fe/H] is an upper limit to the intrinsic
spread.  If significant abundance errors or an abundance gradient exist
within the sample, the estimated intrinsic dispersion, corrected
for these effects, should decrease, 
making the disagreement with the field star sample larger.

Two related papers have attempted to address this issue.
\citet*{GK00} have presented an analysis of
the \citet{E93} data sorted by distance, i.e., separated into two groups divided by
the distance boundary of 30 pc from the sun. The
investigation clearly demonstrates that the more distant sample differs
significantly from the nearby sample in kinematics, metallicity distribution,
and, most important, age-metallicity relation. Though the more distant sample
covers the same range in age as the nearby stars, the average [Fe/H] is
lower in the mean at each age, exhibiting a gradual increase in [Fe/H] with
time. In sharp contrast, the nearby sample shows a rapid rise
in [Fe/H] among the older stars before leveling off among the younger stars.
\citet{GK00} conclude that this difference reflects
the bias in the sample of \citet{E93} toward stars with lower than average [Fe/H],
a bias most easily demonstrated in a distance-sorted sample because of the
need to go to a larger volume to obtain a uniform distribution in [Fe/H],
the point noted above.

The weak point in the discussion, however, is the implication that if an
unbiased sample had been chosen the dispersion in [Fe/H] would reduce
to a value comparable to that of the clusters and HII regions, assuming no 
additional external effects are in operation. \citet{E93} understood the issue
and attempted to correct for it by applying a metallicity-based volume
correction. The adjustment produced only a minor reduction in the dispersion
with age. Thus, either the corrections were grossly inadequate or an 
additional source must be found for the dispersion. It should be
noted that in the three of the four primary
analyses of the age-metallicity relation to date, \citet{TW80}, \citet{ME91}, 
and \citet{E93}, the stars with ages below 2 Gyr exhibited the smallest
dispersion in [Fe/H], while the stars between 4 and 10 Gyr are approximately
constant. The size of the average dispersion, however, has grown larger 
with each subsequent analysis. The one exception to this trend is the 
chromospheric-age-based analysis by \citet{RP00} of 552 field F and G dwarfs.
Using an age scale tied to the system of \citet{E93}, \citet{RP00} derive
an age-metallicity relation with a typical scatter of 0.13 dex, virtually
identical to that found by \citet{TW80}, corroborating the cluster result
of \citet{TAT97} and the claim by \citet{GK00} that some subtle form of selection
bias dominates the \citet{E93} sample. A level of concern, however, is
raised by the fact that the age-metallicity relation derived by \citet{RP00}
shows a linear growth over time, the same result found by \citet{E93} using
a supposedly biased sample. This simple trend is contradicted by 
the analysis of the nearby
sample of \citet{E93} in \citet{GK00} and the analysis by \citet{ME91}
of the unbiased sample of \citet{TW80} using isochrones of the same 
vintage as used by \citet{E93}. Moreover, it leads to a metallicity for
the recently-formed stars in the disk which is comparable to the Hyades,
in contradiction with both young clusters and HII region abundances.
Thus, though we have come full circle in
our view of a well-defined change in the mean metallicity of the disk over
time, the exact shape of the curve remains controversial.

If the disk stars with [Fe/H] below --0.3 are not a statistical
fluke in an otherwise homogeneous solar neighborhood, what are their
origins? The discussion of \citet{GK00} suggests that
part of the problem may arise from contamination of the sample by stars
of the thick disk, irrespective of its origin. This option, however, is
eliminated by the second paper in the pair, the kinematic analysis of
the \citet{E93} data by \citet{QG00}, where it is found that the
velocity dispersion of the disk rises among stars in the age range of 0 to
3 Gyr before leveling off between 3 and 9 Gyrs. The thick disk
makes its appearance in the form of an abrupt increase in velocity dispersion
among stars approximately 10 Gyrs and older. What is important about this
result is that it identifies the thick disk as the product of a specific
event within a narrow range of time. The stars which populate this event
are readily identified in the sample of \citet{E93}; they enter the analysis almost
exclusively based upon their abundance and typically have [Fe/H] = --0.5
or less. If correct, these stars and this event have no relevance to the
origin of the stars between 2 and 9 Gyrs of age with [Fe/H] between
--0.2 and --0.5, the source of the excess dispersion in [Fe/H] relative
to the clusters and the HII regions.

The purpose of this paper is to test the possibility that the cluster data
and the field star data are, in fact, compatible and that the results of
\citet{E93} reflect a combination of both selection bias and stellar orbital
diffusion. In this scenario, the homogeneity of the  cluster 
sample over at least half the
lifetime of the disk is maintained by the lack of a significant
gradient coupled with a cluster dissolution timescale shorter than the
diffusion timescale within the galactic plane. In contrast, the field
stars can be identified at any age. Orbital diffusion of stars formed from 
beyond the break in [Fe/H] at $R_{GC}$ = 10 kpc, coupled with the
age-metallicity relation, readily explains the
inhomogeneity found within the solar neighborhood.

Section 2 will outline the approach to modelling the effects of diffusion
and test the basic program by comparison to previous work, Sec. 3 details
the impact of cluster destruction, Sec. 4 combines the kinematic and
cluster effects with the chemical history of the disk to model
the cumulative sample found within the solar neighborhood, and Sec. 5
discusses the implications of the simulations.

\section{The Model: Background}

The fact that the kinematics of stars within the galactic disk in the
neighborhood of the sun change with time of formation over the life of the
disk is unquestioned. Decades of work have repeatedly shown that the
dispersions in $U$, $V$, and $W$, the stellar orbital velocity components
relative to the local standard of rest,  increase as the mean age of the
stellar sample increases. The primary points of contention are linked to
the details:

(a) What are the exact trends in the dispersions with time?

(b) What physical mechanisms/interactions are the source of the velocity
variation?

(c) What are the relative contributions of the intrinsic dispersion in the
gas from which the stars form and the long term heating of the galactic
disk to the combined velocity trend?

(d) Does the average galactocentric distance of a star's orbit undergo
significant changes over time, irrespective of the mechanism which leads
to the growth in the velocity dispersion?

Clearly, all of these issues are coupled and the extent to which one accepts
or rejects the plausibility of orbital diffusion over extensive ranges
in galactocentric distance, point (d) above, is dependent upon how one
answers (a) through (c). 

The approach adopted here is that developed
by \citet{WI77}, but expanded upon and applied over the last 
two decades by Wielen and his
coworkers \citep{WF85, WF88, WI92, JFW96, WFD96}; readers are 
encouraged to investigate these
papers, particularly \citet{WI77}, for detailed discussions of the underlying
derivations. In short form, stars are assumed to orbit an axially symmetric
galactic potential along paths which are the superposition of a small ellipse
and a circle. The epicyclic approximation \citep{LI59} has the 
ellipse centered upon a
point, the local standard of rest, which moves at constant speed in a circle
about the galactic center. The star moves about the ellipse whose size and
shape are defined by the degree of non-circularity inherent in the galactic
orbit due to the deviations of the star's velocity from that of the local
standard of rest. 

The fact that the observed velocity dispersions in all
three velocity components relative to the local standard of rest 
grow with the mean
age of the sample implies that components of the
potential must exist beyond
the combination of an axially symmetric galactic potential and the effects
of radial variations in the potential. Some additional acceleration process
or processes must exist to perturb the orbits over time. Since star-star
interactions are hopelessly inefficient within the disk, most attempts
to resolve the issue have been constructed about stellar interactions with
larger scale fluctuations in the potential, specifically molecular 
clouds \citep{SS53} or density waves \citep{BW67}. Since their initial 
introduction into the issue, both
forms of the solution have been tested and come up short in their ability
to generate an effect of the required size, though \citet{JE92} has demonstrated
that a combination of some form of both processes may do the trick.

The implicit assumption of the diffusion approach is that while knowing the
exact nature of the mechanism is a valuable check on the reality of the
perturbations, from a statistical standpoint it is a detail which isn't
crucial to modelling stellar kinematics. The acceleration process can
be approximated by a sequence of stochastic perturbations of short
duration; the perturbation changes the velocity of the star relative to
the local standard of rest, but not the position. While the average change
in velocity is zero, the average of the sum of the squares of the changes
in velocity grows with time. The rate of growth of the dispersion with time
is generated by a diffusion coefficient which may take any form with time.
The mathematical form of the coefficient is defined empirically; the product
of the perturbations over time must reproduce the observed changes in 
velocity dispersion over time. The link between the diffusion coefficient
and the velocity dispersions is made by calculating the differences in position
and velocity over time between the stellar distribution with diffusion and in
the absence of diffusion using a set of linearized differential equations which
include the effects of random perturbations in the three velocity components.

In his original study, \citet{WI77} investigated three functional forms for 
the diffusion coefficient: (a) constant with time, (b) constant with time
but inversely dependent upon the peculiar velocity of the star, and (c) 
inversely
proportional to the peculiar velocity of the star and, on average, declining
exponentially with time. Within the large uncertainties imposed by the
observational data, all three functions with an appropriate choice of 
coefficients could reproduce the observations. In general, the dispersions
grew at the slowest pace for case (b), while changing most rapidly among the
oldest stars for case (c). The qualitative conclusion was same
for all three cases. In reproducing the trend in velocity dispersion with
age via stochastic perturbations, within 1 to 2 Gyr the average star will have
its orbit altered so much that any attempt to backtrack the star's orbit
to its origin within the Galaxy using its current position and kinematics
is futile. Quantitatively, all three cases showed significant changes in
velocity early on, but case (b) implied a levelling off of the growth in
the dispersion as the sample aged. 

Subsequent work has focused on a
variety of issues including the size of the diffusion in the radial direction
from the galactic center which implied a significant change in both the
size of the epicyclic ellipse and the instantaneous galactocentric distance
of the center of the ellipse. The importance of the latter issue is that 
analysis of the solar neighborhood sample is often based upon the assumption
that calculation of the mean orbital distance of a star using the instantaneous
velocities provides a direct estimate of the galactocentric location at the
time of formation \citep[see, e.g.,][]{E93}. If radial diffusion is as 
large as Wielen's
approach implies, alternate means of determining the location of origin, such
as combining an assumed abundance gradient 
with age \citep{WFD96}, must be employed.

Because the size of the diffusion coefficient is determined empirically by
optimizing the match of the model to the observed velocity dispersions over
time, much of the criticism of Wielen's work has centered on the claim
that the derived observational trend with age is exaggerated. Both \citet{FR91}
and \citet{QG00} have analyzed the data of \citet{E93} and concluded
that the velocity dispersions only grow with increasing age to 3 Gyr, then
the effect saturates. If true, the diffusion process would be irrelevant
beyond this point and any form of radial diffusion would be limited to the
range imposed by a 3 Gyr time frame.

The criticisms of Wielen's approach are weak on two counts. First,
though the models with constant diffusion over time do exhibit significant
growth in the velocity dispersions among the oldest stars, the model with
a time-constant but velocity-dependent coefficient leads to a rapid rise
in velocity dispersion early on, followed by a gradual
flattening of the curve at increasing age. Despite this declining effect,
radial diffusion remains significant for this model. Second, the claim
that the velocity trend with age remains flat beyond 3 Gyr is not supported
by other analyses of the solar neighborhood, both prior to \citet{E93} and, most
recently, by \citet{FU00} of {\it Hipparcos} data. This question
is often framed in terms of the power-law relationship 
between the total velocity
dispersion and time. The original analysis by \citet{WI77} implied a dispersion
which varied as $t^{0.5}$. A more recent study by \citet*{BDB00} has derived
a power near 0.33, but \citet{FU00} have shown that the discrepancy with
0.5 is mild
and that all the modern data sets are consistent within the errors with
an increasing velocity dispersion over time at a rate similar to that
originally derived by \citet{WI77}. Further resolution of this question may
have to wait for even larger samples of reliable kinematic information.

\subsection{The Model: Galactic Parameters}
Though our primary interest is with the stars within 2 to 3 kpc of the sun, we
have constructed our disk distribution over a galactocentric range 
from 2 to 20 kpc, with the sun located at 8.5 kpc. Clearly, the
reliability of the analysis declines with increasing distance from the sun,
but the effects of diffusion may lead to drifts over distances of a few
kpc, necessitating at least an approximate attempt at including the effects
caused by stars and clusters well beyond the area of interest. The standard 
model begins with 1000 point masses assigned coordinates in a cylindrical
frame centered on the galactic center. The longitudinal coordinate in the plane,
$\theta$, is assigned randomly between  0 \arcdeg and 360 \arcdeg. The radial
coordinate is also assigned in random fashion, but weighted to ensure that
the radial surface density of points produces an exponential variation with
galactocentric distance, typical of what is observed for spiral galaxies and
consistent with the parameters for the luminosity profile of the Milky Way.
For a simple overview of the properties of the Milky Way and galaxies in
general,
the reader is referred to the summary in Chapter 1 of \citet{BT87}; a more
extensive discussion can be found in \citet{BM98}. 

As a simplistic approximation, the rotation curve of the galaxy is assumed to
be flat over the range from 2 to 20 kpc, implying that the Oort constants must
be set such that A = -B.  The value adopted for the rotation curve is $V_{rot}$
= 220 km/sec.  The sample of stars is initially assumed to form 
with an average velocity of
0 relative to the local standard of rest (purely circular orbital velocity about
the galactic center) at each galactocentric distance, but individual stars have
velocities in all three components which are distributed in gaussian fashion
about the mean. The initial gaussian dispersions at $R_{GC}$ = 8.5 kpc 
are selected to resemble the velocity
dispersions found for recently formed stars in the solar neighborhood.
Away from the sun, the velocity dispersion in the $W$ component is assumed
to follow an exponential trend with galactocentric distance with an e-folding
length equal to twice that of the disk \citep{BO93}. Given the relationship between the Oort
constants set by the flat rotation curve, the initial velocity dispersion 
in $U$ becomes a scaled value (1.82) of $\sigma_W$ and the dispersion in
$V$ is a scaled value (0.71) of the dispersion in $U$.

Given $U$, $V$, $W$, and $R$, the galactocentric distance, 
the star's initial epicyclic orbit is derived. Each stars' motion is followed
for two distinct cases:
(a) assuming no perturbations by outside forces and (b) assuming that 
perturbations
are administered to all three velocity components with randomly varying 
strength at regular time intervals ($10^7$ yrs). The size of the 
perturbations is chosen
to ensure a cumulative gaussian distribution at a given time, with the 
dispersion in the gaussian fixed
by the value of the diffusion coefficient. As discussed earlier, the diffusion
coefficient may be defined as a constant, as a time-varying function, or as
a time-varying function of the velocity of the star. Whichever function is 
adopted, the diffusion parameters are selected to reproduce the empirically
defined trends in $\sigma$ over time; for this reason, the timescale between
perturbations is not a factor and has been chosen to ensure a statistically
smooth cumulative alteration in the velocity components on a timescale
comparable to the galactic orbital period of the sun, about 0.2 Gyrs. 
Because of the way the differential
equations are constructed, one can readily calculate the positional
and kinematic differences between 
the unperturbed orbit and the perturbed orbit over time \citep{WI77}.

Though we have modelled the same parameterizations of the diffusion coefficient
as \citet{WI77}, we will only discuss the results for the case where the
coefficient is constant with time but inversely proportional to the velocity
since this case had the smallest long-term diffusion effect on the orbits.
Fig. 1 demonstrates the trend over time for the dispersions in total velocity
and each of the three components averaged from a set of 10 program runs. 
Though the points illustrate the dispersions for the entire sample over all
galactocentric distances, there is no statistically significant difference
compared to the trend if one isolates the stars within 1.5 kpc of the
sun. As expected from previous work, for older stars the velocity dispersion
approaches a relation proportional to $t^{0.5}$.

\subsection{The Model Abundance Gradient and Metallicity Distribution}
Though the Galaxy represents a composite system constructed of multiple
generations of stars, 
to gain some insight into the way in which the disk distribution of stars is
affected by the cumulative perturbations, we first analyze the 
temporal and spatial
evolution of a single generation of stars over a time frame of 10 Gyrs. Our
focus is on the degree to which the galactic abundance gradient is altered
over time, both in shape and dispersion, and the time variation of the
metallicity distribution of stars found within the solar neighborhood, where
we artificially classify the solar neighborhood as the region within 1.5 kpc
of the sun, or the zone between $R_{GC}$ = 7 and 10 kpc.  

For the galactic abundance gradient we select two options. The first case is
the traditional linear gradient with a slope of --0.07 dex/kpc 
\citep[see][and the discussion therein]{TAT97}. The zero-point of this  scale is a major
concern. If we use the clusters inside $R_{GC}$ = 10 kpc, the mean [Fe/H] for
62 clusters in \citet{TAT97} is solar. Given their age distribution, the clusters in
this sample should sample primarily the disk
in the solar neighborhood over the last 2 Gyrs. Unless the 
metallicity of the disk has declined in the recent past, the recently formed
stars in the solar neighborhood should have at least this [Fe/H]. The problem
is that if we extrapolate the gradient to the galactic center, the average
abundance of stars there should be [Fe/H] $\sim$ +0.6. The
stellar population near the galactic center has an extensive but contradictory
history. Originally believed to be super-metal-rich \citep{WR83, RI88, FR90, 
TE91, GF92},
more recent work \citep{MR94, RA00, CSB00} has produced abundances comparable
to solar. The complicating issue is the difficulty of separating true
disk stars from the bulge population but, to date, no evidence exists for any
significant population of super-metal-rich stars near the 
galactic center, bulge or otherwise.

To guarantee that the stars
near the galactic center have a plausible metallicity, we set the mean [Fe/H]
in the solar neighbohood at --0.2, emphasizing again that this is inconsistent
with the data for both young clusters and recently formed stars. The simplest
solution to the contradiction is that the gradient changes slope and 
flattens somewhere within $R_{GC}$ = 10 kpc. Theoretical chemical evolution
models attempting to produce galactic abundance gradients produce such
flattening as a natural consequence of the more rapid evolution of the
inner disk, i.e., the disk develops from the inside out. However, while
evidence exists for flatter gradients among some elements within 4 kpc
of the galactic center, the sizes of the samples are inadequate
to accept this as a definitive observational constraint \citep[see][and the discussion
therein]{HP00}.

The flattening of the abundance gradient is an implicit assumption of the
second
option. For this we assume that there is no gradient within $R_{GC}$ = 10 kpc
or beyond this galactocentric distance. The mean metallicity in the inner
zone is set to [Fe/H] = 0.00, while the mean [Fe/H] of the outer zone is --0.3.
This creates a discontinuity of --0.3 dex at $R_{GC}$ = 10 kpc. Note that the
step function is artificially sharp; it may well be that there are modest
(--0.02 dex/kpc) gradients in both regions and the discontinuity extends over
0.5 kpc. Such distinctions are significant at a level of interest well below
the limits set by the current observational data and minor compared to the
impact of other modelling assumptions. 

In both cases, it is assumed that the stars form at a given galactocentric
distance with a gaussian metallicity distribution and a dispersion of 0.1 dex,
typical of both the cluster sample \citep{TAT97} and the abundance spread among HII 
regions \citep{HW99}, though larger values of 0.2 and 0.3 dex have been tested.
Fig. 2 illustrates the evolution of the abundance gradient 
sampled at $t$ equals 0 Gyrs, after 5 Gyrs of evolution, and after 10 Gyrs for
(a) the linear case and (b) the discontinuity. For the linear case, there is
only weak evidence for change in the slope of the gradient, though the
scatter at a given $R$ clearly grows as stars drift away from their initial
location. From multiple runs of the model, the abundance gradient derived from
data between $R$ = 5 and 15 kpc changes from the initial value of --0.070 to
--0.061 $\pm$ 0.003. In contrast, the dispersion in [Fe/H] grows significantly
from an initial value of 0.10 dex to 0.154 $\pm$ 0.006 dex.

The case for the discontinuity is somewhat more of a challenge to interpret. 
As expected, for regions well away from the discontinuity, the abundance
gradients remain flat with little effect on the dispersion since all the
stars in the surrounding neighborhood have similar abundances. However, as
one studies the disk closer to the break, as the disk ages, diffusion moves
an increasingly larger fraction of stars across the boundary in both
directions. The impact is to create an annular zone where the metallicity
spread approaches a superposition of two gaussians. The greater the passage
of time, the wider the zone where this mixture exists. 

To achieve some sense of what this means for the abundance gradient, we
have derived the gradient in the same way noted above. A linear
relation was fit  to all the data between 5 and 15 kpc. Over this range, the
initial slope was found to be --0.042 $\pm$ 0.002, becoming shallower to
--0.034 $\pm$ 0.002 by 10 Gyrs. Obviously, the slope is strongly dependent
upon the distance range used to define it; if we had chosen 7.5 to 12.5 kpc, the
derived gradient would be similar to the observed value of --0.07. For the
dispersion, the initial spread is typically 0.122 $\pm$ 0.002 dex and increases
only slightly to 0.136 $\pm$ 0.003. The very modest change is readily explained
by the fact that if one uses the entire range between 5 and 15 kpc, the
relative contribution of stars from either side of the discontinuity remains
almost unchanged; the stars mostly shift position within this annulus. 

To get a handle on the importance of diffusion in this model, we need to narrow
the annular range to a more practical value and to move off center from
the breakpoint.  We do this by defining the zone between 7 and 10 kpc as the
solar neighborhood and deriving the metallicity distribution within this
strip. The results are illustrated in Fig. 3, where $a$ and $b$ refer to the
same gradient models as in Fig. 2. For the linear case, the distribution
is initially a gaussian with a slightly broader dispersion than 0.10 dex
because of the range of galactocentric distance from which the
sample is drawn. As time passes, the mean remains approximately [Fe/H] = --0.2,
but the range in metallicity expands as stars from farther afield cross into
the annulus. The lack of a significant asymmetry toward lower [Fe/H], as
observed in the real solar neighborhood, should not be construed as a problem
given that the sample is based upon a single generation of stars with a
well-defined and unchanging metallicity distribution.

In contrast, the discontinuity model starts off with a gaussian distribution
centered on [Fe/H] = 0.0 with a dispersion of 0.1 dex. Since we have chosen
$R_{GC}$ = 10 kpc as the outer boundary, the number of stars from beyond
the discontinuity whose initial orbits temporarily place them within the solar 
neighborhood is negligible. Clearly, this changes as the sample ages to 5 Gyrs
and 10 Gyrs. By the intermediate age a significant tail has developed on the
metal-poor end of the distribution, extending 0.3 dex below the initial lower
edge of the local metallicity distribution. This asymmetry grows stronger in
the oldest sample, producing the telltale profile of two gaussians of 
different weights superimposed. The primary peak in this crudely bimodal
distribution remains relatively unaffected near [Fe/H] = 0.0, while the
secondary peak is hidden beneath the combined sample of the outer disk
distribution and the metal-poor tail of the inner disk distribution.

The obvious conclusion from this simple comparison is that diffusion has a
more dramatic impact upon a discontinuous disk distribution than upon a
linear gradient. For the discontinuity, selection of a sample which
extends across the break will automatically create an apparent gradient
whose slope will depend upon the exact size and location of the
baseline. In both cases, gradients derived assuming a linear
relation remain relatively unchanged over time. The impact of diffusion
is more easily revealed through the metallicity distribution. For 
the linear case, the
sample increases the range in [Fe/H] without significantly altering the 
mean of the distribution. This result is relatively independent of where the
annular sample is selected, though it should be remembered that the flow
into and out of any annulus is not symmetric because the number of
points declines with galactocentric distance. For the discontinuity, the
diffusion of stars from the outer disk creates an extended tail of metal-poor
stars well below the mean [Fe/H] of the inner disk. The size of this tail
and its growth rate are strongly dependent upon the exact location of the
annular strip defined as the solar neighborhood relative to the boundary
of the discontinuity.

\section{The Model: Star Clusters}
Though the analysis above provides a straightforward indication of the
effect of orbital diffusion upon stars, before we construct a 
more realistic model
of the disk with multiple generations of stars, it is important  to test if
the stellar results are representative of what we expect for star clusters.
The issue arises because the presence of the discontinuity appears solely
within the cluster sample \citep{TAT97}, in part because this is the only stellar
sample of statistically significant size which has both reliable distances
and metallicities for objects well away from the solar neighborhood in
both directions. If our
first-order approximation to reality for field stars is correct, diffusion
should gradually soften the edges of the discontinuity until it becomes
almost indistinguishable from a linear gradient. Moreover, in the solar
neighborhood, the sample population should become a mixture of both inner
and outer disk objects, making the definition of the break difficult,
if not impossible. Why, then, do the clusters define such a sharp break?

\citet{TAT97} noted the possibility that the perturbation of an object might be
dependent upon its mass; star clusters with masses thousands of times larger
than a single star will be negligibly impacted as they orbit the disk
in comparison to the effect felt by single field stars. Such an explanation
is very dependent upon the origin of the perturbations, something which
remains a relative unknown. A simpler solution is provided by the
empirical fact that star clusters evaporate over time. 

Assume star clusters form
with the same efficiency and metallicity distribution as the field stars
with galactocentric distance and are perturbed in their orbits to the
same degree as the field stars. All things being equal, the data for
the star clusters should produce the same results as seen in Figs. 2 and 3.
However, as clusters orbit the galactic center, in addition to the
perturbations
they feel along their orbits, tidal forces combined with the perturbations will
also work to remove stars from the gravitational potential well of the 
cluster. The timescale for the evaporation will depend upon where the cluster
is located within the disk, its orbit, the mass of the cluster, and its
radius. Fortunately, we do not need a specific model for the galaxy or
a typical cluster to model the evaporation rate; the rate of cluster
destruction can be derived empirically by observing the age distribution
of a large sample of open clusters over a significant range of the galactic
disk. Such an analysis has been carried out by \citet{JP94}, who find that
the best fit to the data is a superposition of two exponentials. The galactic
models have been rerun with the additional constraint that random points are
removed at a rate consistent with the cluster age distribution. The 
cluster analog of Fig. 2 is presented in Fig. 4. It should be noted that
the 10 Gyr sample is the sum of 10 individual runs of the model. For an
individual run, the typical number of clusters which remain after 10 Gyrs
ranges from 0 to 3, leading to a statistically uninformative plot. By
summing over 10 models, one can get a more valuable sense of how the
overall distribution is affected by location.

The change in the both samples between 0 and 5 Gyrs highlights the critical
point. The reduction in cluster numbers is so dramatic that use of
any sample dominated by older clusters leads to an apparent linear
gradient, whether the original sample exhibited one or not, as long
as there is a statistically significant difference between the 
mean metallicity of the inner and outer disk. The failure to 
recognize the discontinuity is not the product of its dissolution by
cluster diffusion. As expected, because the diffusion timescale is
longer than the cluster evaporation scale, the percentage of clusters
that survive long enough to drift into the solar neighborhood is small.
This implies that with a representative sample of star clusters covering
all ages, a break in the abundance gradient can survive, even though
it is unrecognizable among the field stars.

Before embarking on the final phase of this layered approach, a few 
points regarding the cluster models should be noted. First, the
cluster age distribution is not a true representation of what happens
to a given generation of clusters. The number of clusters which survive
after a time, $t$, is the convolution of the cluster formation rate
over $t$ with the cluster destruction rate, functions which will be
dependent upon where the cluster is formed and where its orbit carries
it over time.  By using all available clusters, \citet{JP94} have
washed out modest variations across the disk, but the relatively
smooth variation in the function with $t$ would seem to indicate that
on the global scale, the formation rate of open clusters has not
oscillated dramatically over the lifetime of the disk. Second, in
contrast with the models, the real disk cluster population exhibits
a paucity of open clusters interior to 7 kpc \citep{JP94}. The
majority of clusters older than 2 Gyr are found in the galactic 
anticenter at respectable distances from the plane, a clear indication
that cluster destruction is more efficient inside the solar circle
and a probable indication that the scale height of the outer disk
is larger than the inner disk.
Fortunately, this contradiction between the models and reality does
not negate the implications of the models in Fig. 4. If one were to
increase the cluster destruction rate in the inner disk and/or lower
the rate in the outer disk, the result would be that the numerical
asymmetry in the cluster points on either side of $R$ = 10 kpc would
be reversed, with the likelihood that the majority of the clusters in
the 5 and 10 Gyr samples would now be found in the outer disk. What 
the plots demonstrate, though, is that as long as one has a significant
baseline to measure the gradient on one side of the break, the 
existence of the discontinuity is still detectable. As will be
demonstrated more effectively in the cumulative model, while definition 
of the exact location
of the discontinuity depends upon a mapping of clusters between 9 and 11
kpc, proof of the reality of the break is best supplied by extending
the baselines of the derived gradients to within $R_{GC}$ = 7.5 kpc 
and beyond 15 kpc. 
  
\section{The Model: Cumulative Evolution} 
Unlike the simple model tested above, the real galactic disk represents a 
cumulative sample of stars and clusters, formed and evolved over approximately
10 Gyr. During that time, the mean metallicity of the newly formed stars rises
as the gas fraction in the disk declines and the products of stellar 
nucleosynthesis are returned to the interstellar medium to become part of
future stellar generations. The rate of star formation and chemical evolution
may vary in a complex fashion over both time and location. When coupled with
the potential impact of galactic orbital diffusion, the resulting trend of
metallicity with galactocentric distance will be a strong function of the
age distribution and galactocentric baseline defined by the data sample used
to measure it.

In an effort to gain some insight into the plausible effects of the primary
processes under discussion, the galactic parameters will be defined as simply
as possible. The qualitative impact of deviating from this simple picture will
be discussed in the last section. As in Sec. 3, two options for the galactic 
abundance gradient will be tested, a linear case and a discontinuity. It is 
assumed that the slope/shape of the disk abundance gradient remains unchanged
with time. However, the time dependence of the zero-point of the scale is 
set such that the value
of the metallicity of the disk at $R_{GC}$ = 8.5 kpc, the location of the
sun, follows an age-metallicity relation approximated by the trend found
in \citet{TW80} and corroborated in \citet{ME91}. The star formation rate
is assumed to be a constant as a function of time.

The cumulative distribution of 11,000 stars after 10 Gyr of 
evolution is illustrated
for one run of the model in Fig. 5. Fig. 5a (top) shows the distribution of
the metallicity with galactocentric distance between $R_{GC}$ = 5 and 15
kpc for the linear gradient while Fig. 5b (top) exhibits the comparable
data for the discontinuous model. The lower plots in Fig. 5 show the same
data on a scale between 7 and 10 kpc. What is readily apparent is that at
all galactocentric distance, the range in [Fe/H] is large, despite the fact
that stars at any given location form with a dispersion of only 0.1 dex.
This is due to a combination of both orbital diffusion and the age-metallicity
relation.  Second, both disks exhibit an apparent gradient. The exact value
of the gradient is dependent upon the baseline over which it is measured,
particularly for the discontinuous case. For the linear gradient case, the 
stellar sample shows derived abundance gradients of --0.065, --0.065, 
and --0.058 dex/kpc over the
distance range of 0 to 20 kpc, 5 to 15 kpc, and 7 to 10 kpc, respectively.
Within the uncertainties, these are identical. For the discontinuous model,
the comparable numbers are --0.021, --0.038, and --0.029. Thus, the field
stars will produce an observable gradient on any scale, but the size of the
gradient will be enhanced in the discontinuous model by the inclusion of a
sample skewed in the galactocentric direction of the discontinuity.

A more relevant comparison for the field stars is that of the 
metallicity distribution. The distribution of stars as a function of [Fe/H]
is illustrated in Fig. 6, following a pattern similar to that laid out in Fig. 5.
The upper plots include stars at all distances  for the two cases, while the
lower figures illustrate the respective models using only stars in the
$R_{GC}$ = 7 to 10 kpc range. The cumulative effects of the differences
in the gradient are obvious in the top figures. Because  the linear slope
is assumed to be constant at --0.07 dex/kpc, if the age-metallicity relation
requires a mean [Fe/H] near solar over the last 5 Gyrs, the metallicity
near the Galactic center approaches +0.6, producing a significant fraction
of stars with [Fe/H] above 0.2. In contrast, this metal-rich tail is
virtually absent from the discontinuous model. Note, however, that
the obvious differences between
these two cases disappear if we restrict our data to field stars in the solar
neighborhood, as seen in the lower plots of Fig. 6. The metal-rich 
stars of the inner galaxy are rarely found 
in the galactic suburbs, eliminating the high-metallicity tail. However, the
low-metallicity portion of the distribution is found in both samples, the
product of the age-metallicity relation and modest orbital diffusion in both
cases.

In contrast, the trends for star clusters are much cleaner, as illustrated
in Fig. 7 for the usual cases. In both instances, the observed gradient for
the clusters has the same well-defined pattern as adopted for each model. The
reason is that the samples are invariably dominated by the clusters formed over
the last 2 Gyr. A handful of clusters manage to survive beyond 5 Gyr, producing
an occasional low metallicity outlier at a given distance due to the lower 
[Fe/H] among older clusters and a possibility of significant drift. For the
most part, the sample still reflects the abundance gradient of the most
recently
formed stars. If the slope/shape of the gradient of newly formed stars remains
relatively unchanged over time, the cluster sample offers the best hope of
detailing the gradient over large galactocentric distance, followed
closely by high mass stars with readily defined distance moduli, e.g.,
Cepheids \citep{FC97}. One can also attempt
to search for time variation of the gradient by sorting the cluster sample
by age but, as noted in the last section, it will be statistically difficult
to derive any pattern except the simplest linear slope through the data points,
irrespective of the detailed shape.

We return now to the issue which initially motivated this discussion, 
the dispersion in the age-metallicity relation. If diffusion does not 
occur, and the intrinsic
dispersion of newly forming stars remains a constant with age of the disk,
and we can measure
stellar ages with perfect accuracy, the metallicity dispersion as a 
function of age should remain a constant defined by the dispersion 
found in the solar annulus at the time of formation of the disk. Though the
dispersion in [Fe/H] is assumed to be 0.1 dex, the changing [Fe/H] across
the solar annulus raises the initial value for the linear gradient model to
0.114 dex. Any change in this value is a product of diffusion. The calculated
age-metallicity relation for the cumulative sample is shown in Fig. 8 
for the (a) linear model and (b) the discontinuity. The
error bars around each point show the size of the dispersion in [Fe/H] at a
given age. The trends for both models
are similar: for samples of increasing age between 0 and 2 Gyrs, the 
dispersion grows by about
50 \%  before leveling off for the discontinuous model while increasing slightly
for the linear case for older stellar samples. The typical 
model dispersion in [Fe/H] among stars from the old disk is between
0.15 and 0.17 dex. Taking into account the additional impact of uncertainties
in age and metallicity, this dispersion is compatible with the results of
\citet{TW80}, \citet{ME91}, and \citet{RP00}, but inadequate to meet the
demands of \citet{E93}, emphasizing again the likelihood that the last
sample is affected by a selection bias which creates an anomalously large
dispersion.

\section{Summary and Conclusions}
The primary goal of this investigation has been a  differential comparison 
of the effects
of orbital diffusion triggered by perturbations within the galactic disk on
a sample of field stars and and a sample of star clusters. Whether all stars
initially form within clusters or the majority of stars form within the field
and small groups \citep{AM01}, it is assumed that both populations follow the
same metallicity trend with galactocentric distance and age. The specific 
galactic gradients modelled include a simple linear trend with a constant
slope over the entire disk at all ages and a discontinuous trend, essentially
flat on either side of a 0.3 dex drop near $R_{GC}$ = 10 kpc. For the
cumulative
evolution of the disk over time, clusters are destroyed at a rate consistent
with the observed age distribution of clusters within a few kpc of the sun.

In the single generation comparison, the effect of diffusion for both 
gradients is to dramatically increase the dispersion in [Fe/H] among the 
field stars within the solar neighborhood, arbitrarily defined as a 3 kpc-wide
annulus centered on the sun. Though the slope of the gradient remains 
essentially unchanged for the linear case, diffusion from beyond the solar
circle washes out the edge of the discontinuity and creates an artificial
gradient among the stars in the solar neighborhood. The exact size of the
gradient depends upon the galactocentric baseline used. For the cluster
sample, the dominant effect is cluster evaporation. If 
all clusters survived, the
expectation is that the trends noted for the field stars would be reproduced
by the clusters. However, cluster evaporation reduces the sample by such an
extreme amount that the probability of a cluster diffusing across the
boundary and surviving for more than a few Gyrs is extremely small.

Despite the artificially restricted nature of the chemical evolution of the
cumulative disk, the patterns noted above are enhanced and the differences
between the models diminished. For the field stars, the time variation of
[Fe/H] expands the range of metallicity included in the final sample well
beyond that imposed by the abundance gradient. Though a significant
difference in the [Fe/H] range is apparent over the entire disk because of
the need to extrapolate the linear gradient to high [Fe/H] near the 
galactic center, within the solar circle typical of the pool used to select
field stars for analysis, the metallicity distributions of the two gradient
models become virtually indistinguishable and bear an amazing resemblence
to the real field star distribution.  Both models produce an apparent
gradient; the diffusion destroys any possibility of identifying the
discontinuity among the field stars.

In sharp contrast, the cluster sample provides an ideal means of distinguishing
between the two cases. The fundamental reason is that cluster evaporation
guarantees that any significant, random sample of clusters will be dominated
by the clusters formed over the last 2 Gyrs. On this timescale, the effects
of diffusion are mild and the sample retains the gradient which existed at
the time of formation. Unless the gradient undergoes serious evolution on a
2 Gyr timescale, open clusters offer the best stellar option for detailing
the recent galactic abundance gradient over large galactocentric distances.
One could attempt to identify the gradient in the past by isolating older
open clusters. However, what the models clearly demonstrate is that, like
the field stars, even if a significant statistical sample can be collected,
diffusion will wipe out any discontinuity, making a distinction between
a real linear gradient and a diffused discontinuity difficult, if not
impossible. Moreover, if one uses the traditional sample of fewer than
20 old disk clusters, any attempt to distinguish between the two cases
or even evaluate a potential variation in the gradient with time is
statistically futile. One will basically define the minimal case, a linear
relation, whose slope may bear little relation to the true trend with
the galaxy at the time of formation. Thus, claims regarding the consistency
of the abundance gradient found among old clusters with that defined by
the young disk are more likely an indication 
of the inadequacy of the sample
than a measure of a fundamental characteristic of galactic evolution. 

Given the above, two issues come to mind. First, are the models representative
of the real disk or are the features merely artifacts of an overly simplistic
description of the disk? In an absolute sense, the models undoubtedly produce
only a mild reflection of reality, particularly in the cumulative case. 
If the source of the perturbations were known, one could vary the size of the
effect with time and galactocentric distance. It is improbable that the
abundance gradient maintains a constant shape over 10 Gyrs worth of evolution;
chemical evolution is likely to occur at a higher pace closer to the galactic 
center, describing an age-metallicity relation which differs from that found
near the sun. Unfortunately, attempts to detect an age trend by defining the
gradient using classes of objects with different mean ages (old clusters
vs. young, HII regions vs. planetary nebulae) are often swamped by the
statistical uncertainty in defining the slope within each category. 
Additionally, chemical evolution models may be constructed which produce 
abundance gradients which grow steeper or shallower with time, making it
difficult to know which trend is appropriate for our model \citep{HP00}.
As a general rule, the effect of diffusion is to make a gradient slightly
shallower over time while smoothing out discontinuities over galactocentric
distance. If the abundance gradient grows shallower over time because of the
delayed increase in the metallicity of the outer disk relative to the inner,
the primary impact will be to smooth the transition between the flatter gradient
in the inner disk and the steeper gradient away from the center
while making the outer gradient somewhat weaker.

More directly, we know from observation that the dissolution
of star clusters occurs at a faster rate within $R_{GC}$ = 7.5 kpc. The
model cluster distribution maintains the same approximate profile with
galactocentric distance over 10 Gyrs, leading to a final sample heavily
weighted toward the center of the Galaxy. With the exception of the last point,
which is predominantly an observational constraint addressed below, most
of the modelling issues which arise dominate on scales beyond the solar circle.
Over the primary galactocentric range of interest, 7 to 10 kpc, the errors
in the absolute sense do not significantly alter the differential comparisons
of the models. There is little question that the net effect of diffusion
is to destroy the detailed structural information content of the field star
sample over time. The bigger the adopted diffusion coefficient and/or the
larger the time frame, the more likely it is that one will be unable to
distinguish
between a simple linear gradient or a more complex gradient structure. Despite
this uncertainty, a derived linear gradient among the field stars implies a
difference in [Fe/H] of some form between the inner disk and the outer disk.

Second, if the field stars lose their ability to inform us about the
detailed structure of the disk over time, can the cluster population
resolve the issue? Because one has the capability of
reliably identifying clusters by age and isolating the younger portion of the
distribution, the best approach to looking for
structure in the galactic abundance gradient lies with the cluster population.
By default, a random sample of clusters will invariably be dominated by the
younger portion of the disk, which explains why the sample of \citet{TAT97} was capable
of identifying the discontinuity. More important is that fact that if a
sudden transition in slope is found for [Fe/H] as a function of galactocentric
distance using a statistically significant sample of younger clusters, it 
is highly probable that it is real. (A
similar conclusion could be postulated for field stars if one could reliably
define age and metallicity for a respectable sample of field stars out to a
distance of 2.5 kpc on either side of the solar circle.) Adjustments
to account for the simplistic
assumptions of the model noted above tend to have, at worst, a weak impact
on the local gradient or, if included, would weaken the appearance of a
discontinuity. As an example, if the star formation rate isn't a constant
over time but was higher in the earlier history of the disk, this
would increase the probability of stars of lower [Fe/H] diffusing into
the solar circle and expanding the observed range of [Fe/H] locally, while
placing more metal-rich stars and clusters beyond $R_{GC}$ = 10 kpc, turning
a potential sharp gradient into a shallow one. Even for the clusters, the
fraction that survive will increase and weaken the feature relative to
the current sample. This will have no impact, however, if one uses only
clusters below 2 Gyrs in age. While it is a straightforward task to destroy
a sharp feature through time evolution or observational error, it is 
equally difficult to create a sharp transition through such means in a 
sample where the feature was initially absent.

We close by noting one additional test of the reality of the feature that
is within the limits of cluster observation. A common criticism of the
discontinuity is the claim that the slope is a product of too small a 
baseline, particularly for the sample on the solar side of the break.
The majority of the clusters studied ($\sim$62) by
\citet{TAT97} lie between 7.5 and 10 kpc
while the small sample beyond the break (14) ranges over 5 kpc. If the break is
real and there is little or no gradient on either side, the critical test
is to study the gradient in the two zones independently, i.e., expand the
sample inside 10 kpc to a distance closer to $R_{GC}$ = 2.5 kpc, and add
more clusters to the 5 kpc range beyond the break. Any analysis which
includes the zone between 8 and 12 kpc will be guaranteed exhibit a measurable
linear slope, whether that is the correct form or not. 

\acknowledgements

The basis of this investigation originated from discussions with Heather
Morrison and Paul Harding of CWRU, whose insights are gratefully acknowledged.
Discussions of ongoing work with Don Garnett were invaluable. We are 
indebted to the referee for prompt and informative feedback to our
original manuscript.
The initial exploration of this problem and the ultimate completion
of the analysis were made possible by the generous support supplied 
by a Goldwater Fellowship to S.C.

\clearpage
\figcaption[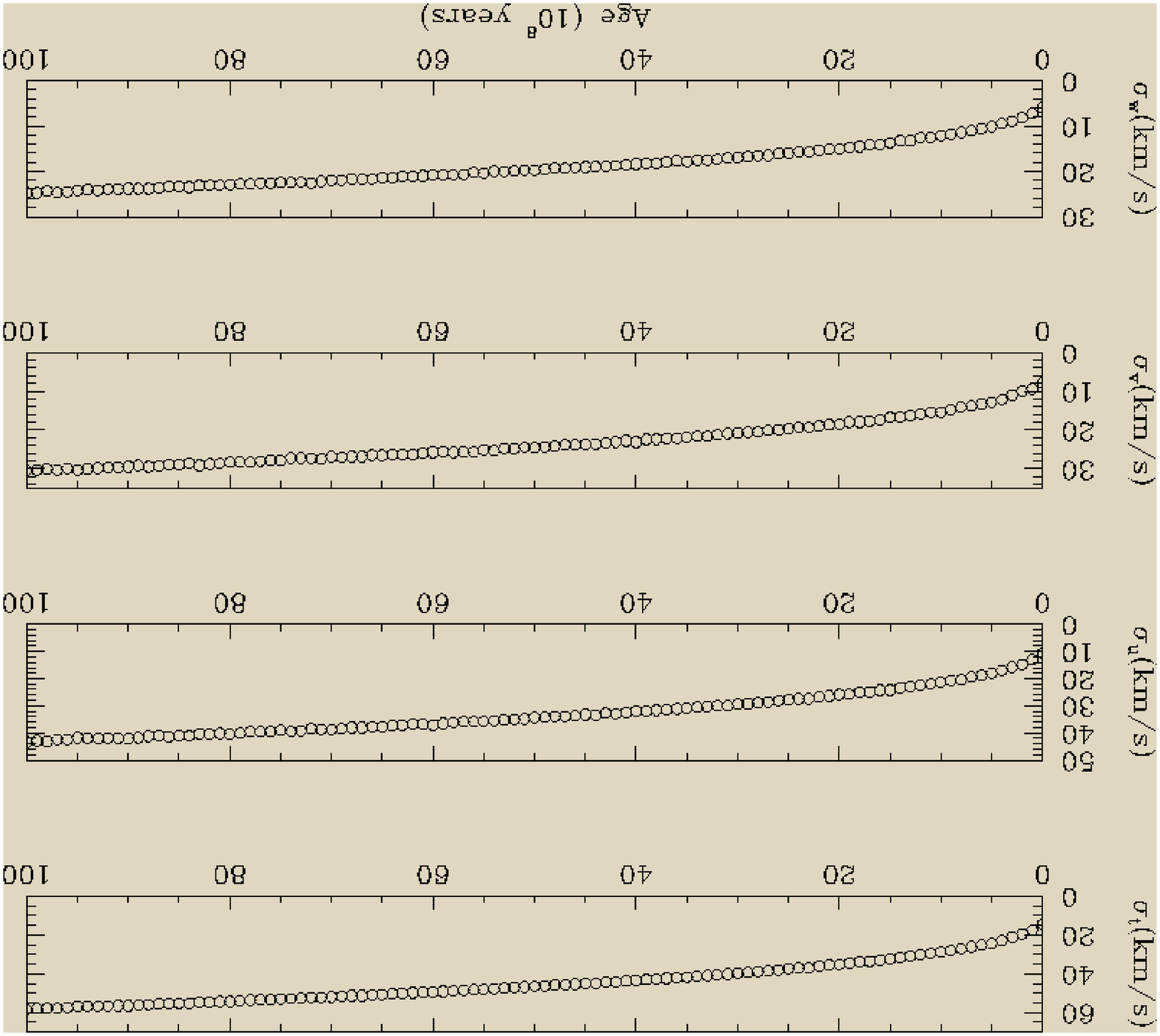]{Velocity dispersions for the individual velocity
components
and the total velocity as a function of time for the model with a
velocity-dependent diffusion coefficient. \label{fig1}}

\figcaption[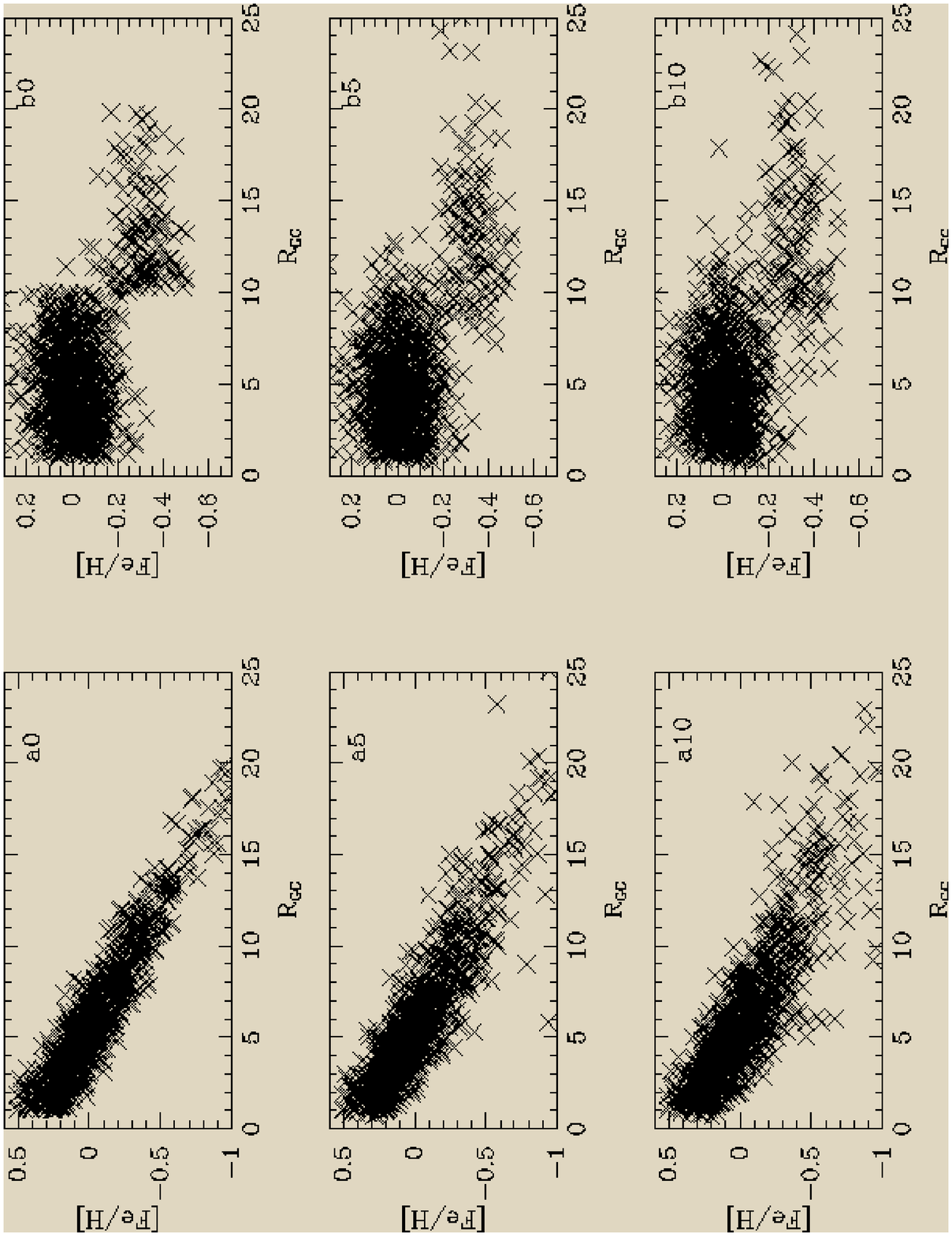]{Evolution of the abudance gradient for (a) the linear case
and (b) the discontinuity. The number after the letter gives the age in Gyrs.
Note that the [Fe/H] ranges differ for the two cases. \label{fig2}}

\figcaption[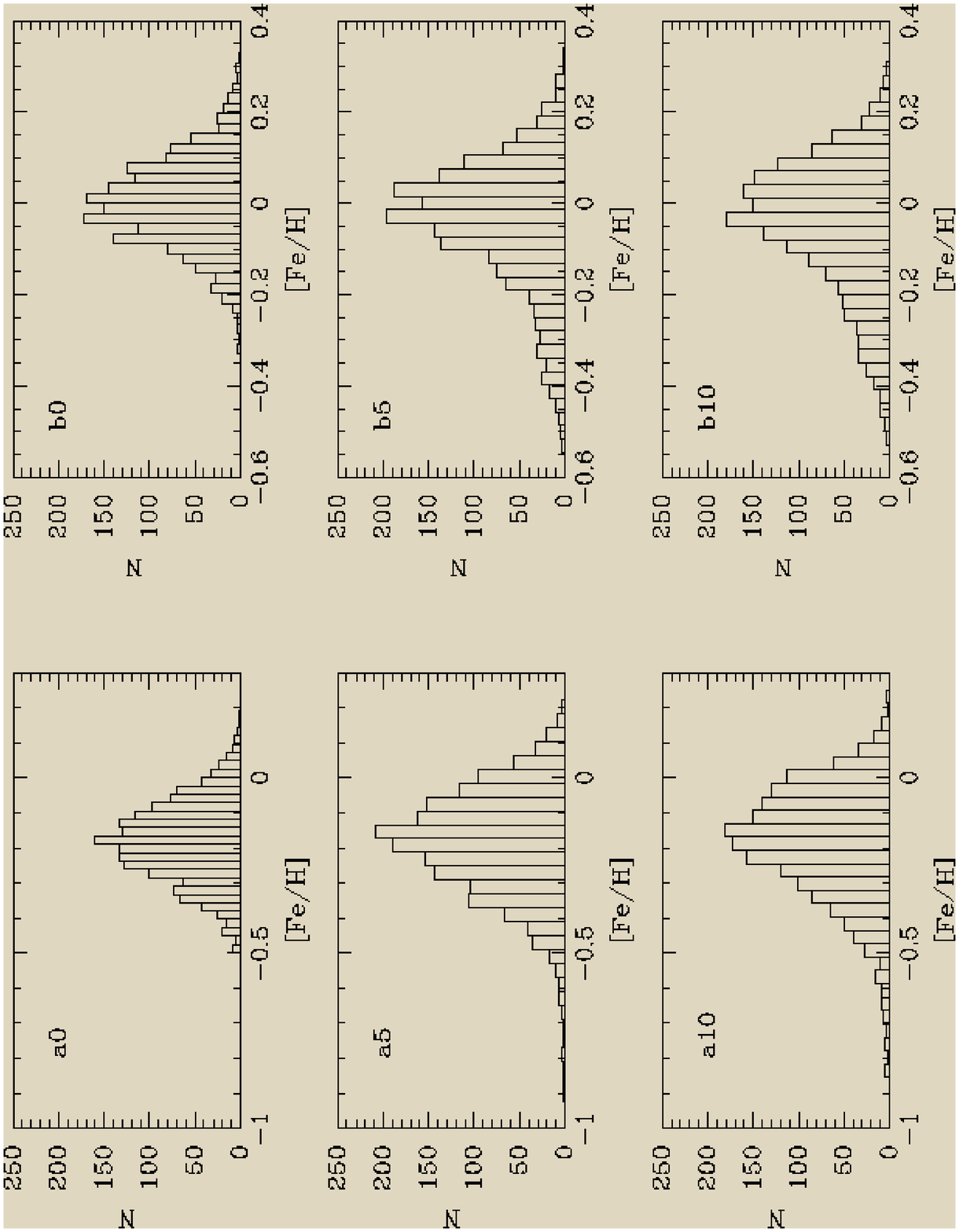]{Metallicity distributions for stars between $R_{GC}$ = 7
and
10 kpc for the same models detailed in Fig. 2. \label{fig3}}

\figcaption[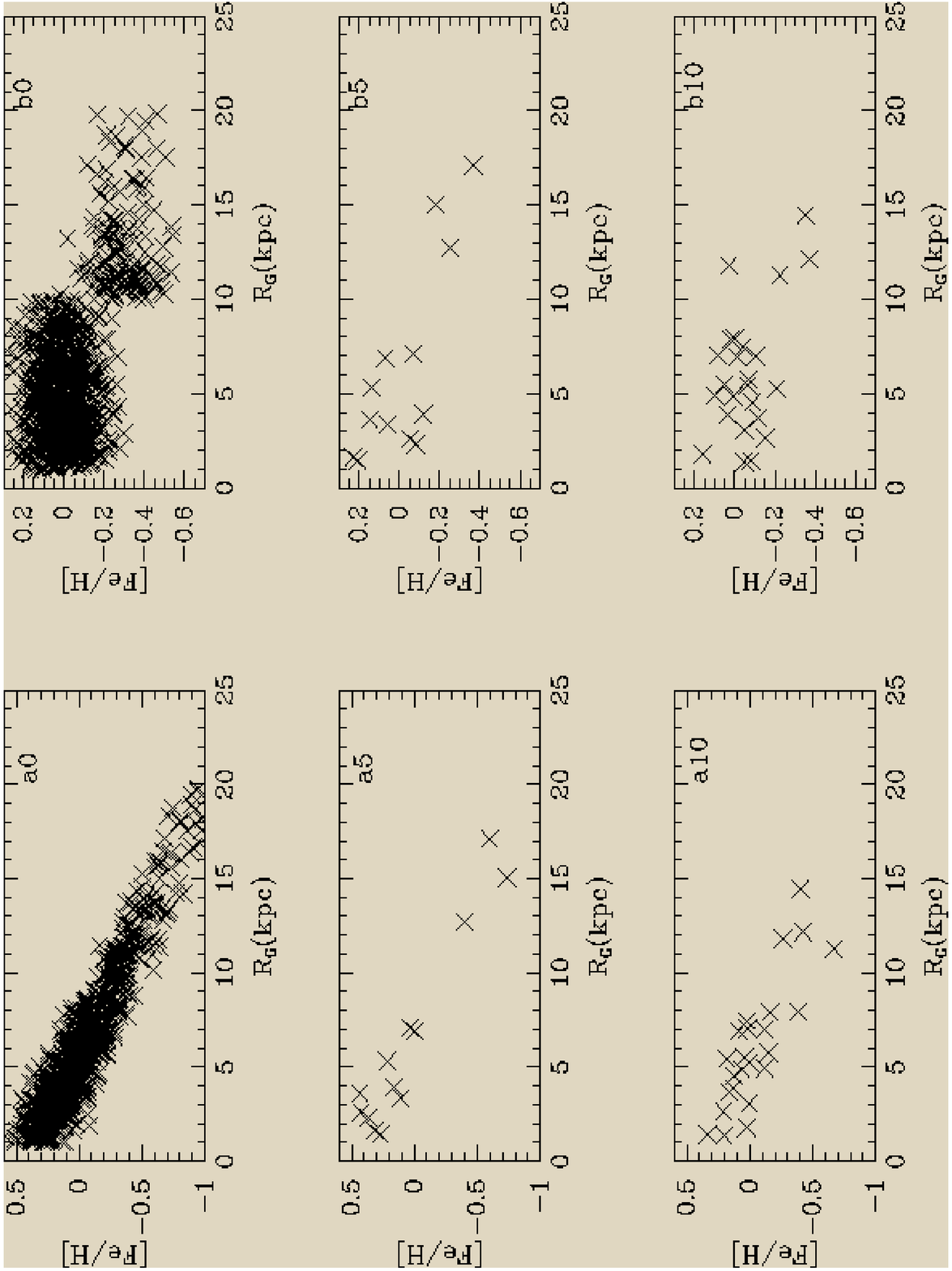]{Same as Fig. 2 but for open clusters. The 10 Gyr sample is
the sum of 10 model runs. \label{fig4}}  

\figcaption[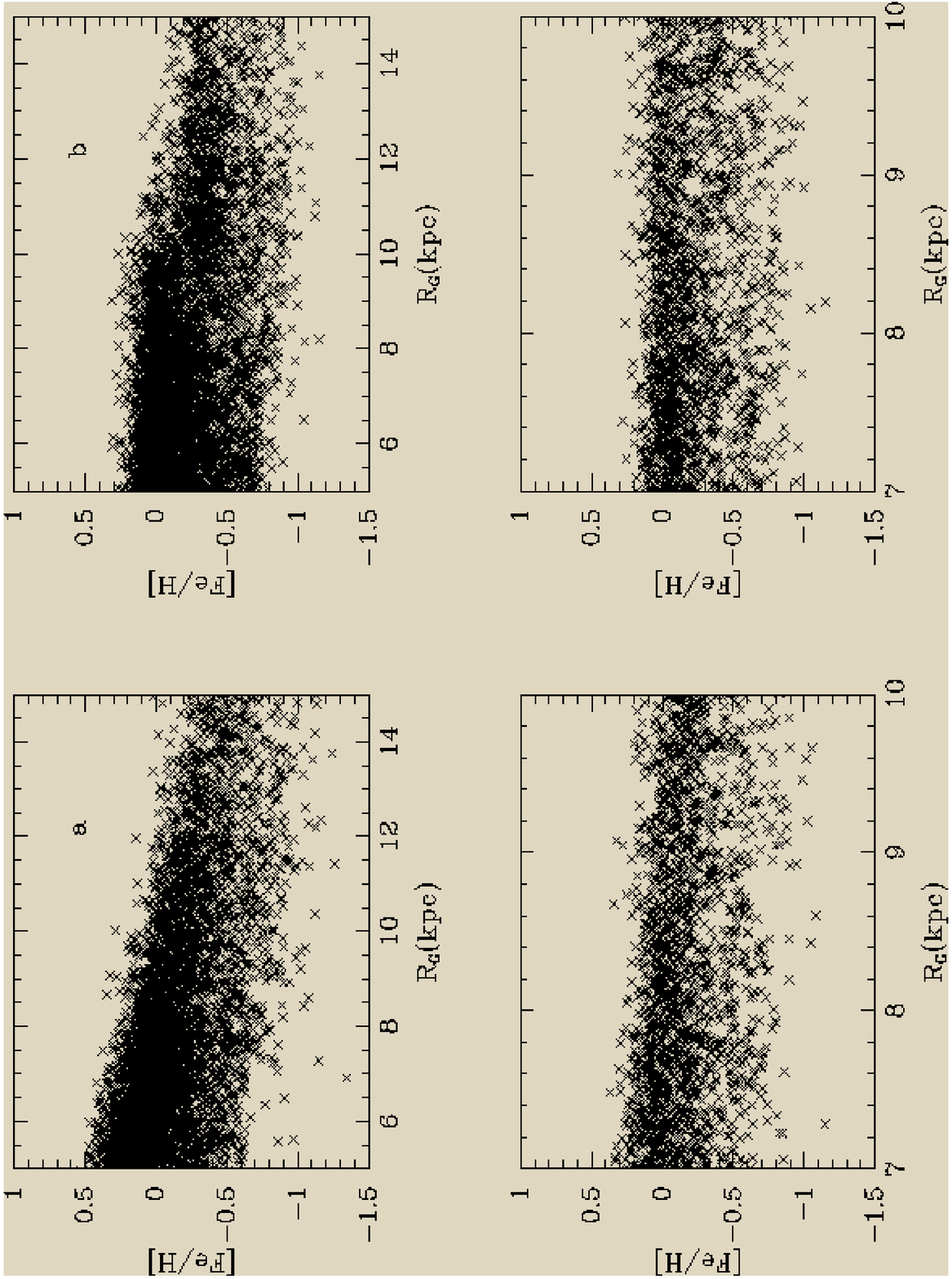]{Metallicity distribution with galactocentric distance for
(a) the linear gradient and (b) the discontinuous disk after 10 Gyr of
cumulative
evolution. The lower figures show an expanded view of the region defined as
the solar neighborhood. \label{fig5}}

\figcaption[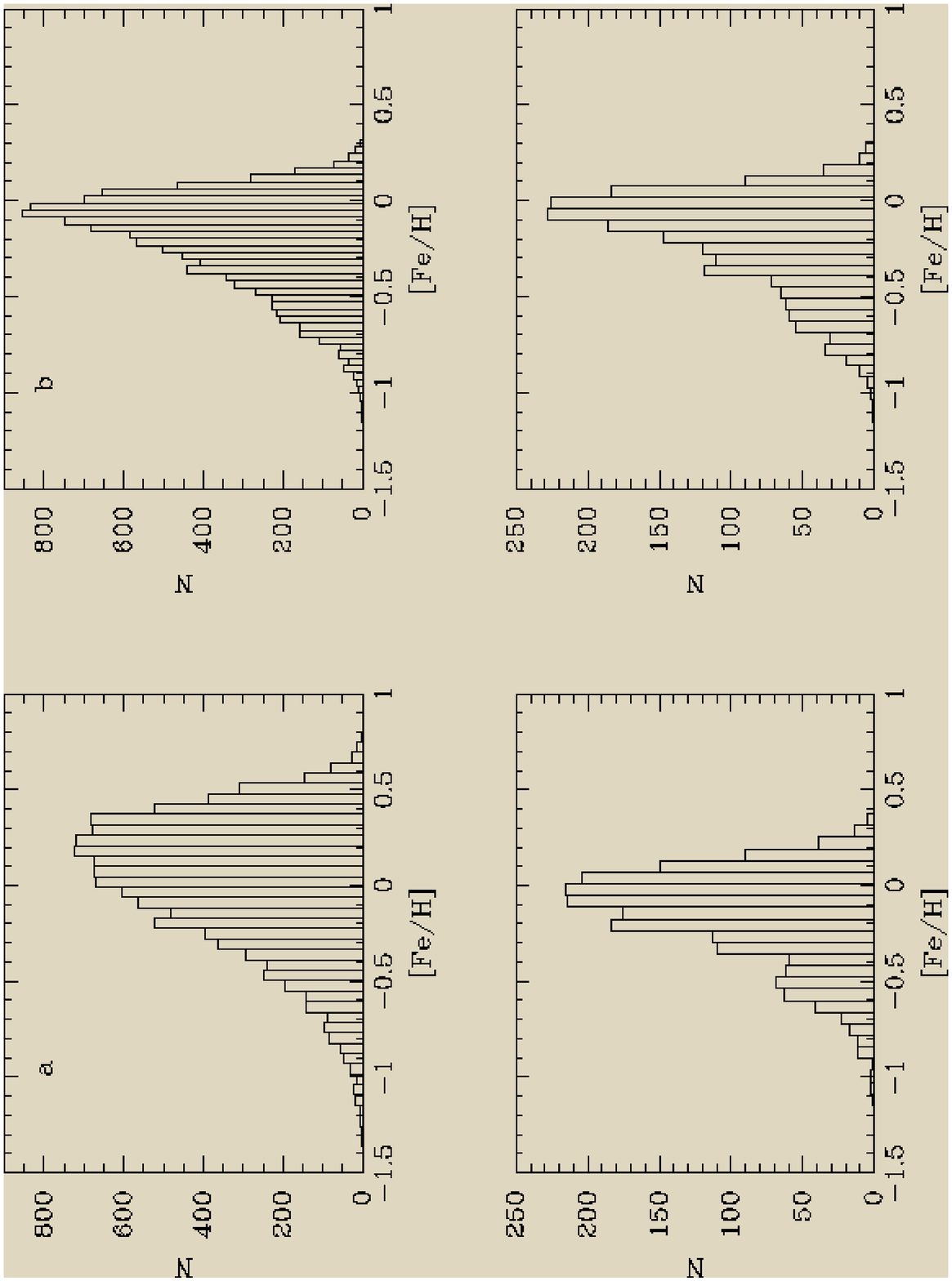]{Metallicity histograms for the same cases as in Fig. 5,
covering the entire range os galactocentric distance in the upper figures and
the solar neighborhood in the lower figures. \label{fig6}}

\figcaption[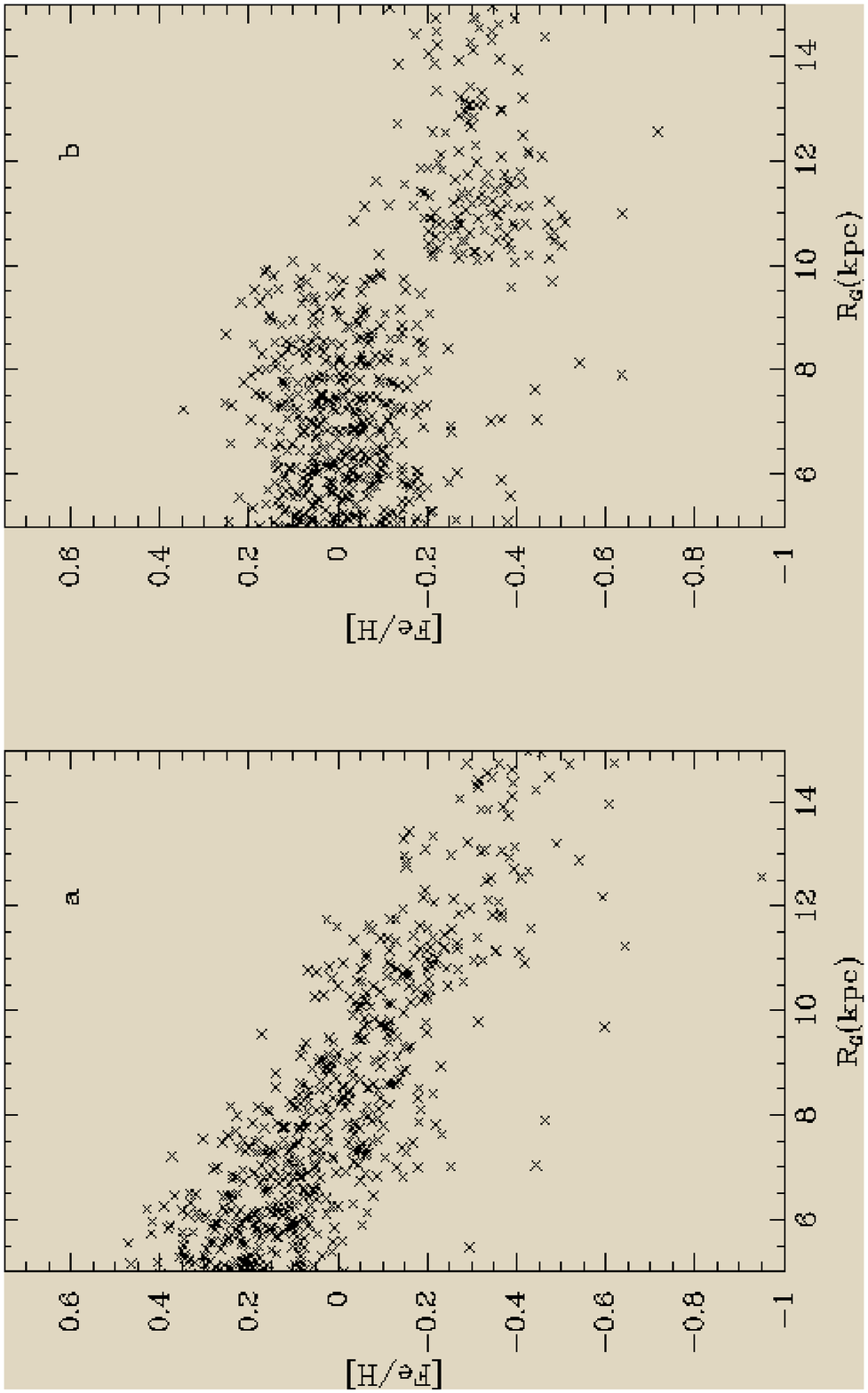]{Metallicity trend with galactocentric distance for
star clusters under the assumption of (a) a linear gradient and (b) a
discontinuous
disk. Sample includes all clusters which form and survive over the 10 Gyr
lifetime of the disk. \label{fig7}}

\figcaption[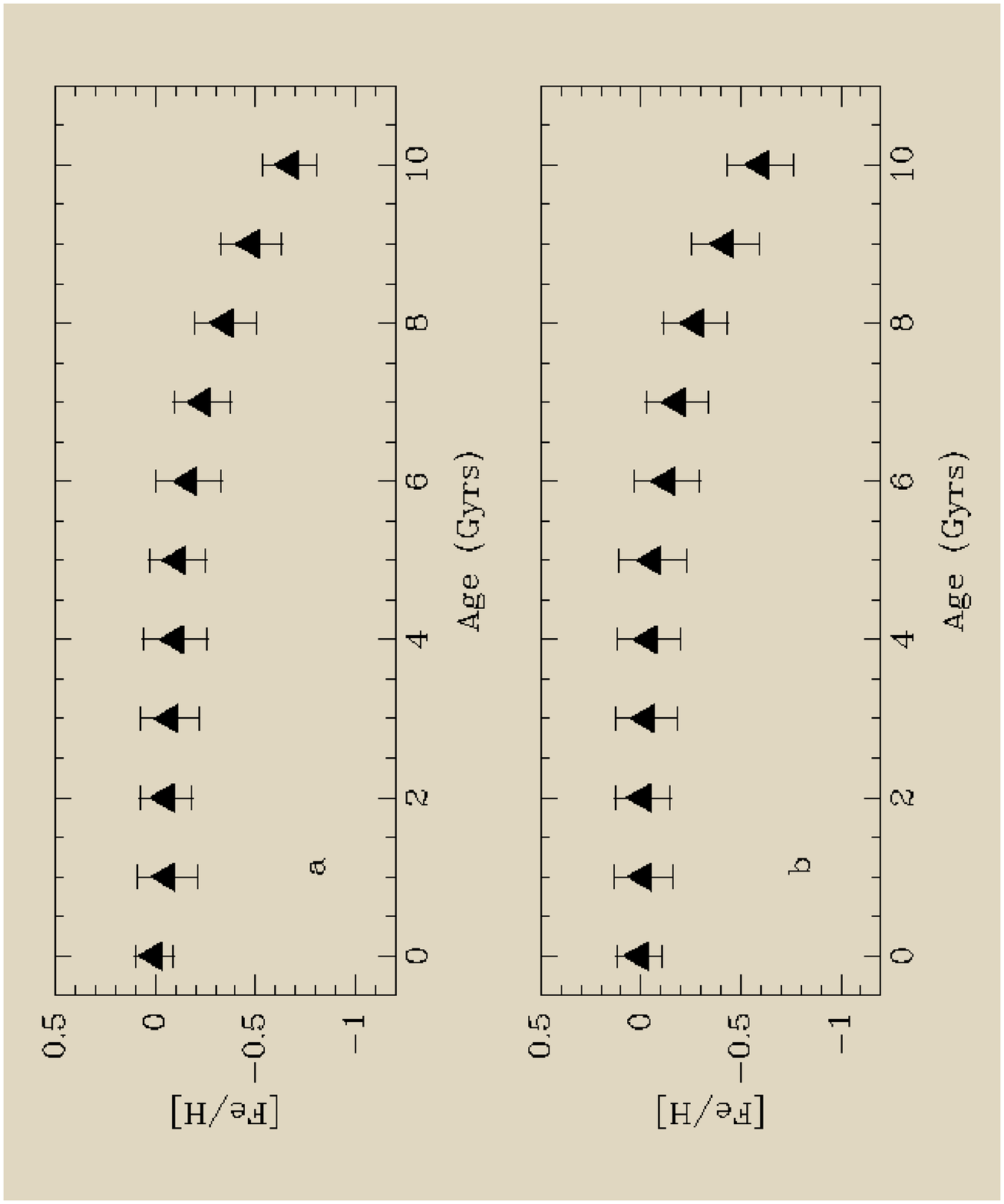]{Age-metallicity relations for the models with (a) 
a discontinuous gradient and (b) a linear gradient for stars in the solar 
neighborhood. \label{fig8}} 
\enddocument